\documentclass[10pt, onecolumn]{article}   	
\usepackage{geometry}                		
\pdfoutput=1                 		
\usepackage{graphicx}
\usepackage{subfigure}				
\usepackage{amssymb}
\usepackage{url}
\title{\bf Statistical Analysis of Dice CAPTCHA Usability}
\author{Darko Brodi\'c$^1$\footnote{(Corresponding author)}, Alessia Amelio$^2$, Ivo R. Draganov$^3$}
\date{\small $^1$ Technical Faculty in Bor, University of Belgrade, Vojske Jugoslavije 12, 19210 Bor, Serbia, \\dbrodic@tfbor.bg.ac.rs\\$^2$Department of Computer Engineering, Modeling, Electronics and Systems, University of Calabria, \\Via P. Bucci Cube 44, 87036 Rende (CS), Italy,\\aamelio@dimes.unical.it\\$^3$Faculty of Telecommunications, Technical University of Sofia,\\ 8 Kl. Ohridski Blvd, Sofia 1000, Bulgaria \\idraganov@tu-sofia.bg}							

\begin{document}
\maketitle

\begin{@twocolumnfalse}

{\bf Abstract}. In this paper the elements of the CAPTCHA usability are analyzed. CAPTCHA, as a time progressive element in computer science, has been under constant interest of ordinary, professional as well as the scientific users of the Internet. The analysis is given based on the usability elements of CAPTCHA which are abbreviated as user-centric approach to the CAPTCHA. To demonstrate it, the specific type of Dice CAPTCHA is used in the experiment. The experiment is conducted on 190 Internet users with different demographic characteristics on laptop and tablet computers. The obtained results are statistically processed.  At the end, the results are compared and conclusion of their use is drawn.
\end{@twocolumnfalse}
\vspace{1cm}

\section{Introduction}
CAPTCHA (Completely Automated Public Turing test to tell Computers and Humans Apart) is a program representing a challenge-answer to the given test, which is used to realize if the solver is an Internet user (human) or a computer program (computer robot). This process includes the following elements: (i) The computer as a server, which generates the CAPTCHA test, (ii) Internet users or computer program which try to correctly solve the given task, (iii) The computer which evaluates the answer to the CAPTCHA in the format Yes/No (correctly/incorrectly solved). Typically, the CAPTCHA task is accustomed to the humans. Hence, there is a greater possibility that humans will solve this task compared to computer robots abbreviated as bots. Hence, the aim of the CAPTCHA program is to differentiate Internet users from bots \cite{[1]}. 

The application of CAPTCHA program is useful in the following areas: (i) Online systems, (ii) The creation of free e-mail accounts, (iii) Online pooling, and (iv) Online system for buying tickets, etc. \cite{[2]}.

Still, the CAPTCHA should fulfill certain elements, such as: (i) The solving of CAPTCHA should not rely on the user's knowledge of certain language, (ii) The solving of CAPTCHA should not depend on the user's age, (iii) CAPTCHA should make an automatic evaluation of the correctness, (iv) The user's privacy should not be violated, and (v) It should be easy for Internet users to be solved unlike bots \cite{[3]}.

The related works on CAPTCHA often employ statistical approaches treating their various aspects. They can be partitioned taking into account their properties in the following areas: (i) Security, (ii) Practicality, and (iii) Usability \cite{[4]}.

Security represents the main concern to the CAPTCHA programmers. It represents a central problem of CAPTCHA, but it is not the only one that is of a great importance.
 
Practicality is connected to the way of creating certain types of CAPTCHA. Again, it has greater concerns of programmers than CAPTCHA users. 

The usability represents the main problem related to the use of the CAPTCHA. Accordingly, it especially concerns the CAPTCHA users. Hence, this study is used to uncover the elements of CAPTCHA usability, which represents the main concern of the Internet users. In this way, an objective analysis of a certain type of CAPTCHA can facilitate better understanding the user-centric relation between computer and man, i.e. CAPTCHA and Internet user which will contribute to innovate and improve CAPTCHA elements to be more accustomed to the Internet users unlike bots.

This paper is organized in the following manner.  Section II presents the CAPTCHA types. Section III describes the experiment. Section IV gives the results of the experiment and discusses them. Section V draws conclusions and points out the direction of future works.

\section{CAPTCHA Types}
All CAPTCHA types can be divided into five typical groups: (i) Text-based CAPTCHA, (ii) Image-based CAPTCHA, (iii) Audio-based CAPTCHA, (iv) Video-based CAPTCHA and (v) Other types of CAPTCHA \cite{[5]}.

\begin{figure}[!b]
\begin{center}
\includegraphics[height=9cm, width=9cm, keepaspectratio]{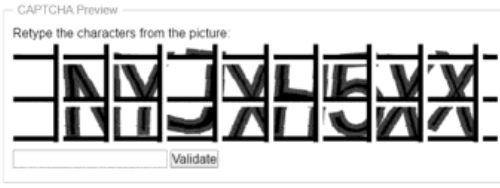}
\caption{An example of the text-based CAPTCHA}
\label{Fig1}
\end{center}
\end{figure}

Text-based CAPTCHA asks the Internet users to input exact combination of the given characters. This type of CAPTCHA is the most widespread one. In order to reduce its vulnerability to bot attacks, many distorted elements are incorporated. Unfortunately, the text-based CAPTCHA can be successfully attacked by bot due to the solid OCR (Optical Character Recognition) programs. Fig. \ref{Fig1} shows an example of the text-based CAPTCHA.

\begin{figure}[!b]
\begin{center}
\includegraphics[height=7.5cm, width=7.5cm, keepaspectratio]{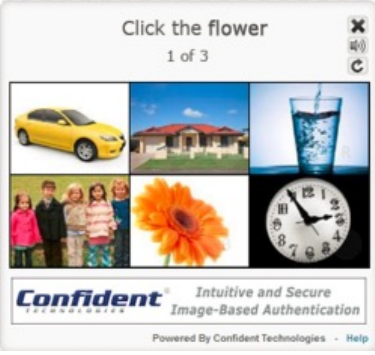}
\caption{An example of image-based CAPTCHA}
\label{Fig2}
\end{center}
\end{figure}

Image-based CAPTCHA is considered as one of the most advanced and safest types of CAPTCHA. It requires from the users to find out a certain image from a list of images and point to it. Due to that, its elements include the image details. It represents a relatively easy task to be solved by Internet users unlike bots. Fig. \ref{Fig2} illustrates an example of the image-based CAPTCHA. 

\begin{figure}[!t]
\begin{center}
\includegraphics[height=9cm, width=9cm, keepaspectratio]{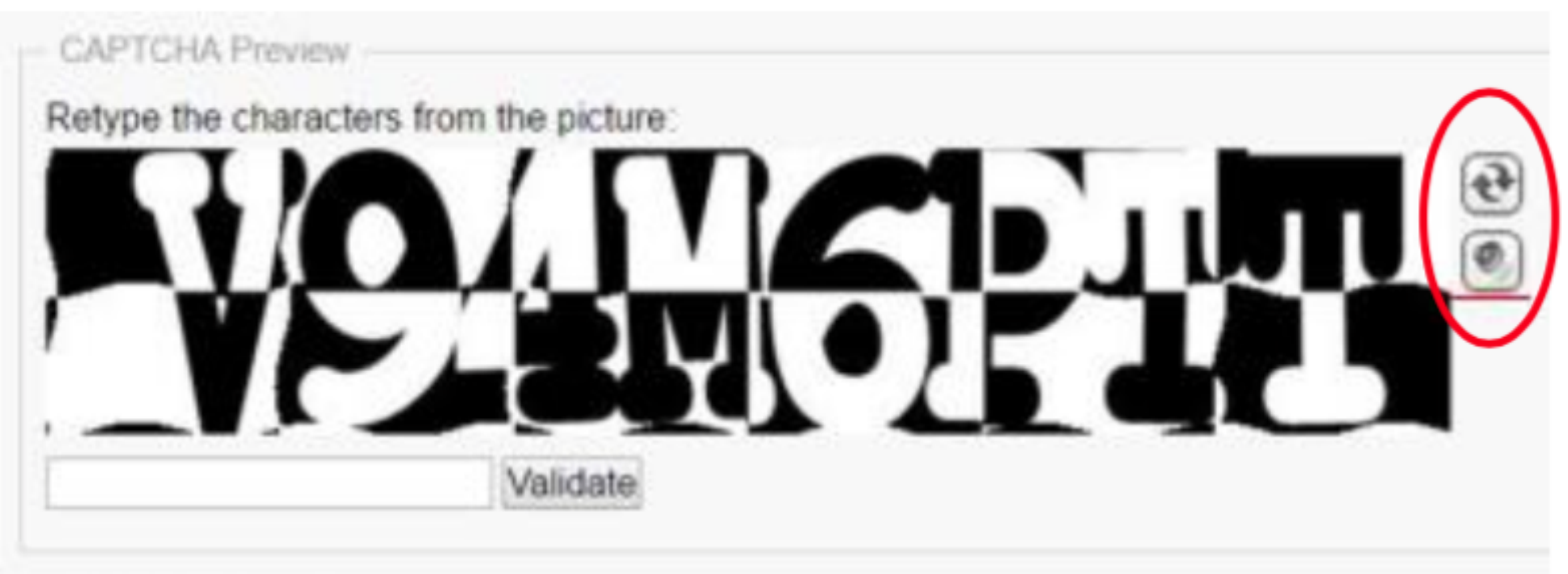}
\caption{An example of the audio-based CAPTCHA}
\label{Fig3}
\end{center}
\end{figure}

Audio-based CAPTCHA includes an Òaudio elementÓ whose purpose is an audio reproduction of characters that the user should have to input in order to solve the CAPTCHA. This type of CAPTCHA is especially designed for the people with disabilities. Unfortunately, the audio-based CAPTCHA is mostly attacked by speech and recognition algorithms in approximately 70\% of cases. Fig. \ref{Fig3} illustrates an example of the audio-based CAPTCHA with Òaudio elementÓ in the top right corner.

\begin{figure}[!t]
\begin{center}
\includegraphics[height=9cm, width=9cm, keepaspectratio]{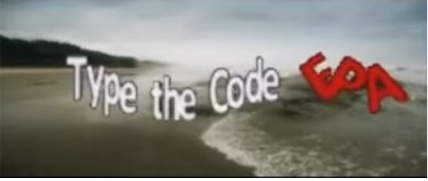}
\caption{An example of the video-based CAPTCHA}
\label{Fig4}
\end{center}
\end{figure}

Video-based CAPTCHA contains text information embedded into the video. Hence, it is a video which includes a passing text given in specific color compared to video background.  The user should recognize the given passing text and type it. The modern OCR programs challenge this task, making this CAPTCHA vulnerable to bot attacks. Fig. \ref{Fig4} illustrates an example of the video-based CAPTCHA.

\begin{figure}[!t]
\begin{center}
\includegraphics[height=9cm, width=9cm, keepaspectratio]{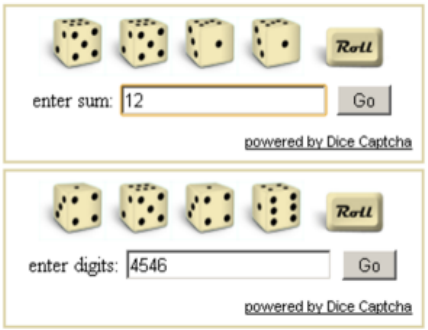}
\caption{An example of other type of CAPTCHA on the Dice CAPTCHA samples (Dice 1 at the top, and Dice 2 at the bottom)}
\label{Fig5}
\end{center}
\end{figure}

Other types of CAPTCHA represent those CAPTCHAs that cannot be part of the previous standardization. Fig. \ref{Fig5} illustrates the examples of such types of CAPTCHA.

\section{Experiment}
The CAPTCHA experiment is conducted on 190 Internet users. It is divided in two different experiments solving two different Dice CAPTCHAs (Dice 1 and 2). The first experiment is based on Dice CAPTCHAs tested on a community of 90 laptop users aged from 29 to 62 years. The laptop used for the experiment is Lenovo B51 with the following characteristics: (i) 15.6" wide screen, (ii) CPU Quad-core 2.4 GHz Celeron, (iii) 4 GB of RAM, (iv) 500 GB of internal memory, and (v) Operating system Microsoft Windows 7. The second experiment is based on Dice CAPTCHAs tested on a community of 100 tablet users aged from 28 to 55 years. The tablet used for the experiment is Lenovo IdeaTab A3000 with the following characteristics: (i) 7" wide screen, (ii) CPU Quad-core 1.2 GHz Cortex-A7, (iii) 1 GB of RAM, (iv) 16 GB of internal memory, and (v) Operating system Android. 

\section{Results and Discussion}
\subsection{Hypotheses}
It is worth noting that the solution time for ``ideal" CAPTCHA should not depend on the age, education and gender differentiation. However, if any CAPTCHA can satisfy these elements, then it doesn't mean that it can be solved quickly and easily. According to previous facts, the following four hypotheses are proposed according to the given demographic characteristics:

Hypothesis 1 (H1) - There exists a statistically significant difference between users' groups (laptop vs. tablet) in average response time to solve the CAPTCHA.

Hypothesis 2 (H2) - There exists a statistically significant difference between age groups in solving the CAPTCHA,

Hypothesis 3 (H3) - The group of Internet users with higher education will have a faster response time in solving the CAPTCHA,

Hypothesis 4 (H4) - There exists a statistically significant difference between gender groups in solving the CAPTCHA.

\subsection{Experimental Results}

The first results of the experiment are given in Tables I-II. These tables give a descriptive analysis of the obtained results for 90 laptops' and 100 tablets' Internet users concerning CAPTCHA Dice 1 and 2. They are obtained by Kolmogorov-Smirnov test, which tests the unknown distribution and checks the normality assumption in the analysis of variance \cite{[6]}.

\begin{figure}[!ht]
\begin{center}
\includegraphics[height=9cm, width=9cm, keepaspectratio]{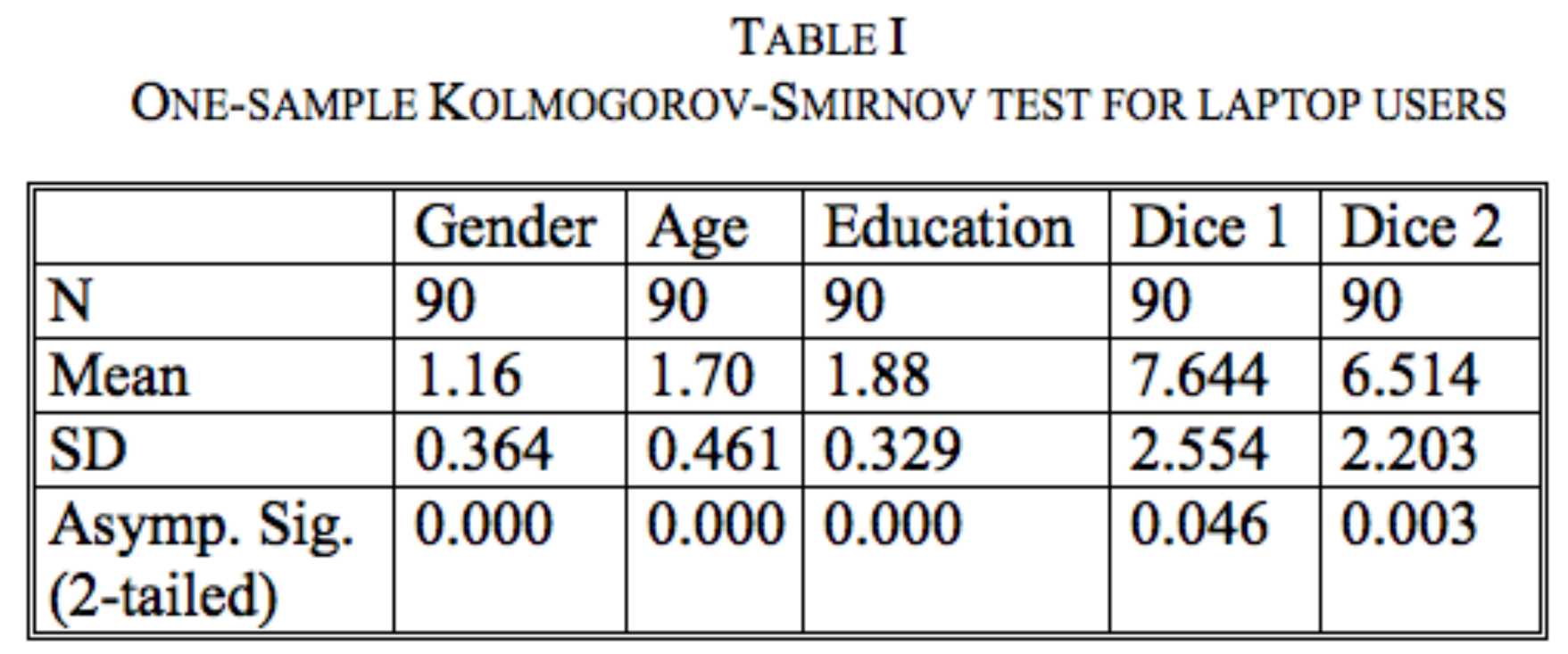}
\end{center}
\end{figure}

The most important information represents the measure Asymp. Sig. (2-tailed). It defines the statistical significance of the analyzed data. Because it is smaller than 0.05, then obtained results are statistically significant. Also, it is worth noting that the average time to solve CAPTCHA Dice 1 is 7.6444, while CAPTCHA Dice 2 is solved in 6.514 seconds by laptop users.  

\begin{figure}[!ht]
\begin{center}
\includegraphics[height=9cm, width=9cm, keepaspectratio]{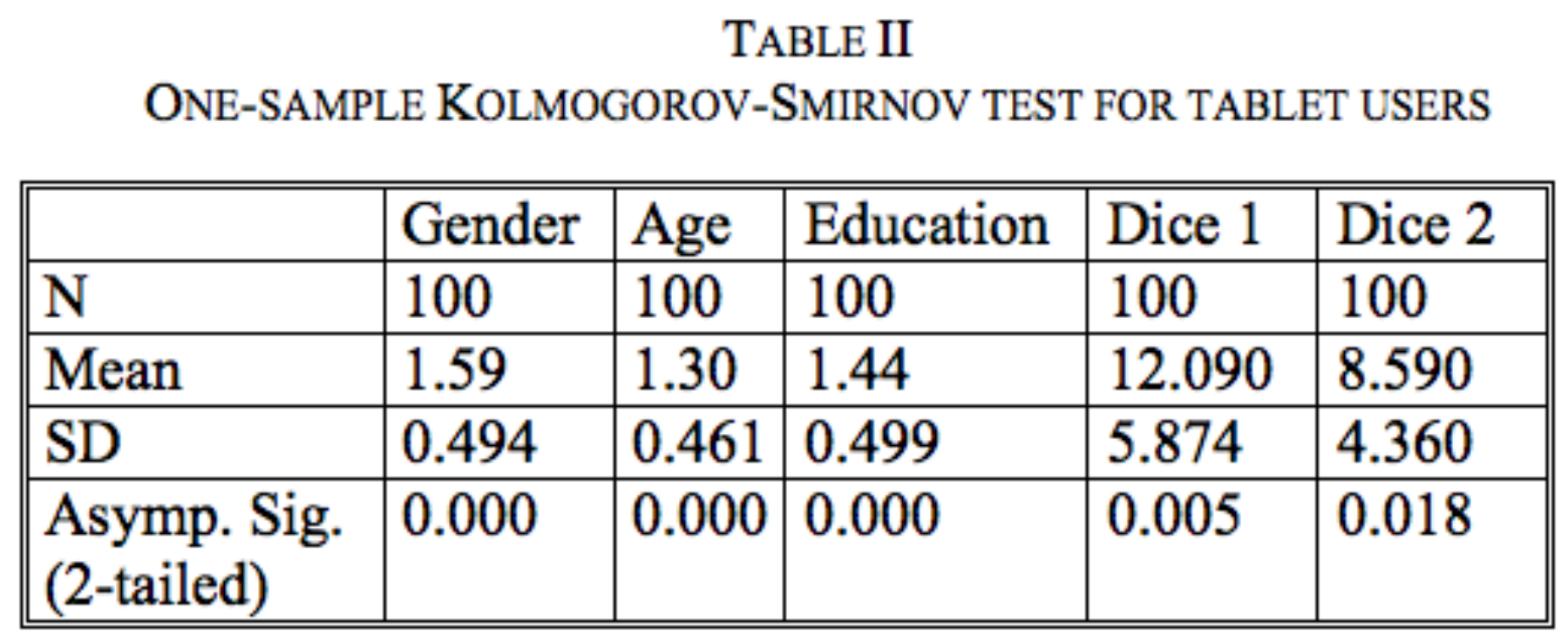}
\end{center}
\end{figure}

From Table II, the measure Asymp. Sig. (2-tailed) is again lower than reference of 0.05, which determines the statistical significance of the analyzed data. Furthermore, it is worth noting that the average time to solve CAPTCHA Dice 1 is 12.090, while CAPTCHA Dice 2 is solved in 8.590 seconds by tablet users.  

From Tables I-II, it is quite clear that exists a statistically significant difference in the response time to solve CAPTCHA Dice1 and Dice 2 between laptop and tablet users' groups. Obviously, Dice CAPTCHA is more convenient to be solved on a laptop than on a tablet computer. It is proved by statistical significant population. Hence, H1 is proved. 

\subsection{Statistical Test}

The Mann-Whitney $U$ test is a non-parametric test which can be used to (dis)prove a null-hypothesis $H_0$ and a research hypothesis $H_1$. Essentially, this test is used to compare differences between two independent groups $N_1$ and $N_2$. To be used, some pre-assumptions should be valid: (i) Input should be composed of two categorical independent groups $N_1$ and $N_2$, (ii) Output should be ordinal or continuous, (iii) There should be no correlation between groups $N_1$ and $N_2$, and (iv) The input variables should not be normally distributed. The Mann-Whitney $U$ test considers 3 important measures: (i) $p$-value, (ii) $U$ value, and (iii) $Z$ value.

The $p$-value is the first crucial measure of this statistical test. Its value can be interpreted as follows: (i) $p < 0.05$ shows a strong evidence against the null-hypothesis. As a consequence, the null-hypothesis of the test is disproved, while research hypothesis $H_1$ is proved, (ii) $p \ge 0.05$ shows a weak evidence against the null-hypothesis of the test. As a consequence, the null-hypothesis of the test is proved, while research hypothesis $H_1$ is disproved. $U$ value is calculated as:

\begin{equation}
U=n_1n_2+ \frac{n_1(n_1+1)}{2}-R_1,
\end{equation}
where $U$ represents the result of the Mann-Whitney $U$ test. Accordingly, $n_1$ is the size of the independent group $N_1$, $n_2$ is the size of the independent group $N_2$, and $R_1$ represents the sum of ranks of group $N_1$. If $U$ value is higher than the critical $U$ value, then the two groups $N_1$ and $N_2$ will have the same score distributions, otherwise the two distributions $N_1$ and $N_2$ will be different in some aspect. Critical value $U$ is important only for small size distributions, where the number of their elements is up to 20. If the group is larger than 20, then $U$ value approaches to normal distribution. In that case, the $Z$ value has importance. It is calculated as:

\begin{equation}
Z=\frac{U-n_1n_2/2}{\sqrt{n_1n_2(n_1+n_2+1)/12}}.
\end{equation}	 	

If the absolute value of $Z$ is lower than 1.96, then the two groups $N_1$ and $N_2$ will have the same score distributions, otherwise the two distributions of $N_1$ and $N_2$ will be dissimilar in some way. Accordingly, if $Z$ is lower than 1.96 research hypothesis is disproved, otherwise it is proved.

\subsection{Analysis of the Results and Discussion}

The results obtained by statistically processing (Mann-Whitney $U$ test) of experimental data for the age characteristic of the laptop/tablet users are given in Table III.

The first relevant measure, which has to be evaluated is Asymp. Sig. (2-tailed). For laptop users as well as for tablet users concerning CAPTCHA Dice 1 and 2 it is higher than 0.05. Accordingly, this analysis is not statistically significant. Hence, $H_2$ is not proved.

The results obtained by statistically processing (Mann-Whitney $U$ test) experimental data for the education demographic characteristic of the laptop and tablet users are given in Table IV.

Again, the measure Asymp. Sig. (2-tailed) is evaluated the first. For laptop users as well as for tablet users concerning CAPTCHA Dice 1 and 2 it is higher than 0.05. Hence, this analysis is not statistically significant. This leads that $H_3$ is not proved.

\begin{figure}[!ht]
\begin{center}
\includegraphics[height=9cm, width=9cm, keepaspectratio]{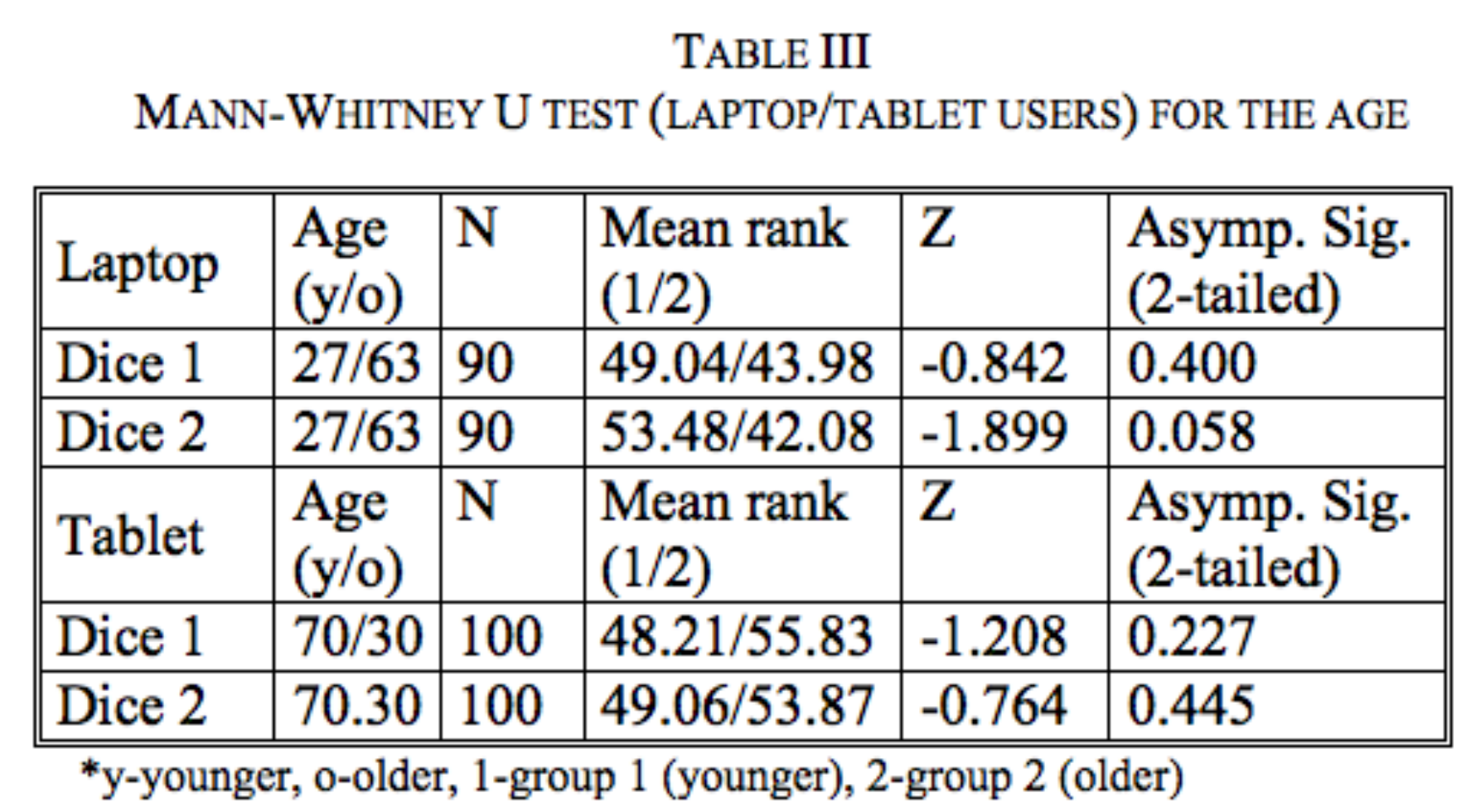}
\end{center}
\end{figure}

\begin{figure}[!ht]
\begin{center}
\includegraphics[height=9cm, width=9cm, keepaspectratio]{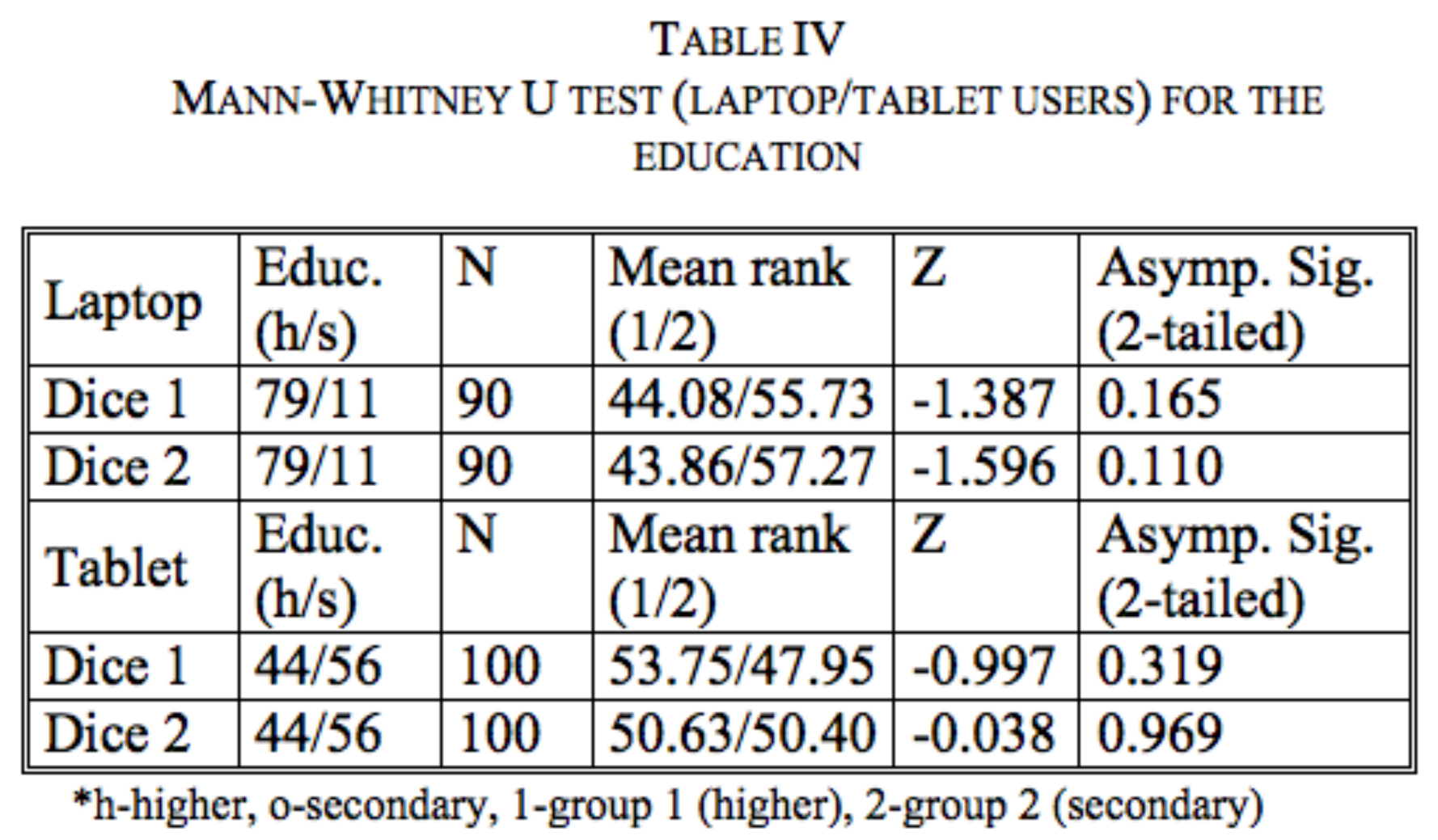}
\end{center}
\end{figure}

The results obtained by statistically processing (Mann-Whitney $U$ test) experimental data for the gender demographic characteristic of the laptop and tablet users are given in Table V. 

\begin{figure}[!ht]
\begin{center}
\includegraphics[height=9cm, width=9cm, keepaspectratio]{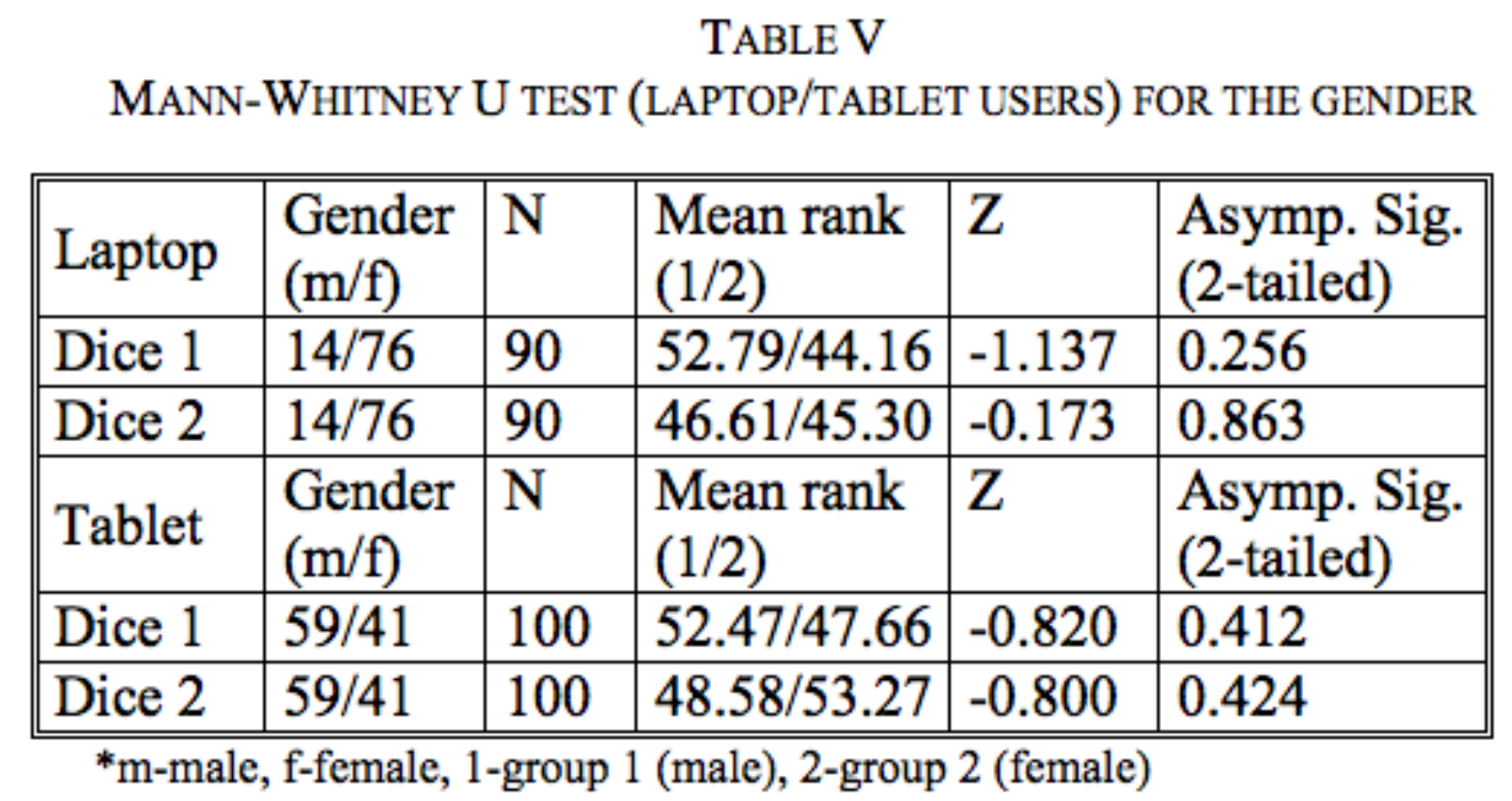}
\end{center}
\end{figure}

From Table V the measure Asymp. Sig. (2-tailed) is again higher than reference value 0.05. Hence, for laptop users as well as for tablet users concerning CAPTCHA Dice 1 and 2 the given analysis is not statistically significant. Accordingly, $H_4$ is not proved.

From the aforementioned, the $H_1$ is only proved, while $H_2$, $H_3$ and $H_4$ are not proved. Because, the postulate of ``ideal" CAPTCHA is to be solved in reasonable time (less than 30 sec. \cite{[5]}), and the solution time should not depend on the age, education and gender differentiation, the Dice CAPTCHA represents a good direction toward creating an ``ideal" CAPTCHA. However, it is worth noting that using CAPTCHA on different computer types should also diminish differences between solution time of certain CAPTCHA. In our case, solution time of Dice CAPTCHA between laptop and tablet users is almost 50\% less in favor of laptop users. Taking into account this information, Dice CAPTCHA is more accustomed to the laptop than tablet Internet users. Hence, Dice CAPTHA can be considered only as the first step in right direction toward creating an ``ideal" CAPTCHA. 

\section{Conclusion}
The paper analyzed the response time of Internet laptop and tablet users in solving the Dice CAPTCHA version 1 and 2. To research the given topic, an experiment was conducted on 190 users. It was divided into two parts: (i) testing of 90 laptop users in solving Dice CAPTCHA 1 and 2, and (ii) testing of 100 tablet users in solving Dice CAPTHA 1 and 2. Then, the obtained results were statistically processed. According to the results, four hypotheses were established, which should be proved or disproved. All hypotheses were closely related to the elements of an ``ideal" CAPTCHA. Using statistical tools, a descriptive statistical analysis and the results of Mann-Whitney $U$ test were used for proving and disproving the given hypotheses. At the end, the $H_1$ hypothesis was only proved, while the other ones were rejected. In spite of the obtained result, which represents the main elements of an ``ideal" CAPTCHA, due to rather different time in solving Dice CAPTCHA between laptop and tablet users, this type of CAPTCHA cannot be used as an example of ``ideal" CAPTCHA. But, because of some overlapping with the characteristics of an ``ideal" CAPTCHA, the Dice CAPTCHA is a good start and a right direction toward creating the real ``ideal" CAPTCHA.  
\\\\\\
{\bf Acknowledgement.}
This study was partially funded by the Grant of the Ministry of Education, Science and Technological Development of the Republic of Serbia, as a part of the project TR33037.

\end{document}